\documentclass[aps,prd,preprint,a4paper,showpacs]{revtex4}
\usepackage{graphicx,natbib}
\usepackage{amsmath}
\newcommand{\dis}{\displaystyle}
\newcommand{\bi}[1]{\mbox{\boldmath ${#1}$}}
\begin{document}
\title{Higgs potential in $S_3$ invariant model \\
for quark/lepton mass and mixing    }
\vspace*{1cm}
\author{T. Teshima}
\email{teshima@isc.chubu.ac.jp}
\affiliation{Department of Natural Science and Mathematics,  Chubu University, 
Kasugai 487-8501, Japan}
\begin{abstract}
We analyzed the $S_3$ invariant Higgs potential with $S_3$ singlet and doublet Higgs. 
We obtained a relation $(|v_1|/|v_2|)^2=-\sin2\phi_2/\sin2\phi_1$  
from this  $S_3$ invariant Higgs potential, where $v_1$, $v_2$ and $\phi_1$, $\phi_2$ 
are vacuum expectation values and phases of  $S_3$ doublet Higgs, respectively. 
This relation could be satisfied exactly by the results  $\dis{|v_1|/|v_2|=0.207}$, $\phi_1=
-74.9^\circ$ and $\phi_2=0.74^\circ$ obtained from the previous our work analyzing the
quark/lepton mass and mixing in $S_3$ invariant Yukawa interaction. 
Furthermore, the relation $v_S \sim v_D=\sqrt{|v_1|^2+|v_2|^2}=174{\rm GeV}$ is obtained and 
then the coupling strength of Higgs to top quark $g_{H_Stt}={m_t/v_S}$ is altered as by a factor
 $\sqrt{2}$ from the standard value.  
Introduced the $S_3$ doublet Higgs, FCNC are produced in tree level.
Predicted branching ratios for rare decays $\mu^-\to e^-e^+e^-$, $K^0_L\to \mu^+\mu^-$  
etc., induced by the FCNC are sufficiently below the present experimental upper bounds.
\end{abstract}
\preprint{CU-TP/12-02}
\maketitle
\section{Introduction} 
The existence of the Higgs bosons which are the origin for Higgs 
mechanism producing the masses of all matter and gauge fields is expected to be discovered in 
LHC \cite{ATLAS, CMS}.   
We analyzed the problem of the origin of quark/lepton mass and mixing using a $S_3$ invariant 
model \cite{TESHIMA1, TESHIMA2},  where quark and lepton flavors and Higgs fields are 
considered to be governed by the $S_3$ symmetry.      
In our model, weak bases of flavors $(u, c)$, $(d, s)$, $(e, \mu)$, $(\nu_e, \nu_\mu)$  
are assumed as $S_3$ doublet  and $t,\ b,\ \tau,\ \nu_\tau$ are $S_3$ singlet and further 
there are assumed $S_3$ doublet Higgs $(\Phi_1, \Phi_2)$ and $S_3$ singlet Higgs $\Phi_S$.  
Constructing the $S_3$ invariant Yukawa interactions, we could explain the quark sector mass hierarchy  
and mixing $V_{CKM}$ including phases of CP violation.  
In the leptonic sector, assuming the see-saw mechanism \cite{SEESAW} with the Majorana masses, 
we could explain the tri-bimaximal-like character \cite{ATMOS, SOLAR, NEUTRINO} of neutrino mixing 
$V_{MNS}$  without imposing any other symmetry restriction than $S_3$ symmetry.    
\par
This minimal $S_3$ extension of the Higgs fields in flavor or generation space could play an important role 
in explaining the quark/lepton mass and mixing \cite{TESHIMA1, TESHIMA2}. 
The ratio of vacuum expectation values $|v_1|, |v_2|$ of Higgs doublet $\Phi_1$, $\Phi_2$ is estimated to 
be not 1 and rather small, $\dis{|v_1|/|v_2|=0.207}$, and this ratio can explain the Cabibbo angle. 
The phases $\phi_1$ and $\phi_2$ of Higgs doublet $\Phi_1$ and $\Phi_2$ are the origins of 
CP violation, and these values are estimated as $\phi_1=-74.9^\circ$ and $\phi_2=0.74^\circ$. 
In this paper, we construct the Higgs potential as $S_3$ invariant, and then investigate
whether this Higgs potential  could satisfy the above results or not. 
\par
We use the most general  $S_3$ invariant Higgs potential adopted by many authors \cite{PAKUBASA, KUBO2, 
DERMAN, BRANCO, KOIDE, BHATTACHARYYA}, assuming a hierarchy among the quartic coupling strengths  of  Higgs  
fields.
A relation  $(|v_1|/|v_2|)^2=-\sin2\phi_2/\sin2\phi_1$ is obtained from the stationary condition in 
our  $S_3$ invariant Higgs potential. 
The relation could be satisfied exactly by the result  $|v_1|/|v_2|=0.207$,  $\phi_1=-74.9^\circ$ and 
$\phi_2=0.74^\circ$. 
Furthermore the relation $v_S \sim v_D=\sqrt{|v_1|^2+|v_2|^2}=174{\rm GeV}$, which should be compared 
to the standard value $v=246{\rm GeV}$,  is obtained and then the coupling strength of Higgs $H_S$ to 
$t$ quark $g_{H_Stt}=m_t/v_S$, is altered as by a factor $\sqrt{2}$ from the standard value.  
Introduced the $S_3$ doublet Higgs, flavor changing neutral current (FCNC) are produced in tree level.
We analyze the branching ratios for rare decays $\mu^-\to e^-e^+e^-$, $K^0_L\to \mu^+\mu^-$  
etc., induced by the FCNC, which are predicted to be sufficiently below the experimental upper bound
obtained from PDG data \cite{PDG10}.      
\section{$S_3$ invariant model for quark/lepton mass and mixing}
First, we review our {$S_3$ invariant model for quark/lepton mass and mixing \cite{TESHIMA1,TESHIMA2}. 
We assumed that the generations of quarks and leptons (charged leptons and Dirac neutrinos) 
and further Higgs fields are the irreducible representation  of $S_3$ symmetry group,     
\begin{equation}
\begin{array}{l}
S_3\ {\rm doublet}\ :\ {\bi f}_D^{L, R}=\!(f_1^{L,R}, f_2^{L,R})^T,\ \ {\bi \Phi}_D=\!(\Phi_1, \Phi_2)^T,\\
S_3\ {\rm singlet}\ :\ f_S^{L, R},\ \Phi_S,\\
\ \ \ \ f_1=u, d, \nu_e, e, \ \ f_2=c, s, \nu_\mu, \mu, \ \ f_S=t, b,  \nu_\tau, \tau.
\end{array}
\end{equation} 
In the $SU(2)_L$ gauge space, 
\begin{align}
SU(2)_L\  {\rm  doublets} \ :\ &  {\bi \Phi}_D=({\bi \Phi}_D^+, {\bi \Phi}_D^0)^T,\   \Phi_S=(\Phi_S^+, 
\Phi_S^0)^T,  \notag  \\ 
&Q_1^L=(u_L, d_L)^T,\  Q_2^L=(c_L, s_L)^T,\  Q_S^L=(t_L, b_L)^T, \notag  \\
& L_1^L=({\nu_e}_L, e_L)^T, L_2^L=({\nu_\mu}_L, \mu_L)^T,\  L_S^L=({\nu_\tau}_L, \tau_L)^T, \\ 
SU(2)_L\  {\rm singlets}\ :\ &  d_1^{R}=d_{R},\  d_2^{R}=s_{R}, \ d_S^{R}=b_{R},\  u_1^R=u_R, u_2^R=c_R,\ 
u_S^R=t_R,  \notag \\
& l_1^R=e_{R},\  l_2^R=\mu_{R},\  l_S^R=\tau_{R},\  \nu_1^R={\nu_e}_R,\  \nu_2^R={\nu_\mu}_R,\  
\nu_S^R={\nu_\tau}_R.   \notag
\end{align} 
As a standard model, Yukawa interaction for Dirac mass of flavors is expressed as 
\begin{equation}
-{\cal L}_D^{f}=\dis{\sum_{i,j,k=1,2,S}}[\Gamma_{ijk}^{d}\overline{Q_i^L}\Phi_j d_k^R+\Gamma_{ijk}^{u}
\overline{Q_i^L}\epsilon \Phi_j^* u_k^R+
\Gamma_{ijk}^{l}\overline{L_i^L}\Phi_j l_k^R+\Gamma_{ijk}^{\nu}
\overline{L_i^L}\epsilon \Phi_j^* \nu_k^R
]+ h.c.,
\end{equation} 
where $\Gamma^{f}_{ijk}$ are interaction strengths and $\epsilon$ is the $2\times2$ 
antisymmetric tensor in the $SU(2)_L$ gauge space. 
In our model, we assumed that the Yukawa interaction Eq. (3) is $S_3$ invariant. 
Under our present analysis considering the effects caused from the neutral Higgs $\Phi_S^0$ 
and ${\bi \Phi}_D^0$,  the Yukawa interactions are expressed as
\begin{eqnarray}
&-{\cal L}_D^{d,l}=\Gamma_S^{d,l}\overline{f^L_S}f^R_S\Phi^0_S+\Gamma_{D1}^{d,l}
{\overline{{\bi f}^L_D}}
{\bi f}^R_D\Phi^0_S+\Gamma_{D2}^{d,l}[(\overline{f_1^L}f_2^R+\overline{f_2^L}f_1^R)\Phi_1^0+
(\overline{f_1^L}f_1^R-\overline{f_2^L}f_2^R)\Phi_2^0]\nonumber\\
&+\Gamma_{D3}^{d,l}(\overline{{\bi f}_D^L}{\bi \Phi}_D^0f_S^R+\overline{f_S^L}{{\bi \Phi}_D^0}^T
{\bi f}_D^R)+h.c., \notag\\
&\hspace{3cm}{\rm for\ down{\mbox -}type\ quark\ and\ charged\ lepton,} \\
&-{\cal L}_D^{u,\nu}=\Gamma_S^{u,\nu}\overline{f^L_S}f^R_S{\Phi^0_S}^*+\Gamma_{D1}^{u,\nu}
{\overline{\bi f}^L_D}{\bi f}^R_D{\Phi^{0}_S}^*+\Gamma_{D2}^{u,\nu}[(\overline{f_1^L}f_2^R+
\overline{f_2^L}f_1^R){\Phi_1^{0}}^*+
(\overline{f_1^L}f_1^R-\overline{f_2^L}f_2^R){\Phi_2^{0}}^*]\nonumber\\
&+\Gamma_{D3}^{u,\nu}(\overline{{\bi f}_D^L}{{\bi \Phi}_D^{0}}^*f_S^R+\overline{f_S^L}
{{{\bi \Phi}_D^{0}}^*}^T{\bi f}_D^R)+h.c. \notag\\
&\hspace{3cm}{\rm for\ up{\mbox -}type\ quark\ and\ Dirac\ neutrino.}\nonumber
\end{eqnarray}
These mass Lagrangians are almost similar  to those in literature analyzing the $S_3$ invariant model 
\cite{PAKUBASA, KUBO1, KUBO2, MONDRAGON}, where 
$\Gamma_{D3}^{d,l}(\overline{{\bi f}_D^L}{{\bi \Phi}_D^{0}}f_S^R+\overline{f_S^L}
{{{\bi \Phi}_D^{0}}}^T{\bi f}_D^R)$ and 
$\Gamma_{D3}^{u,\nu}(\overline{{\bi f}_D^L}{{\bi \Phi}_D^{0}}^*f_S^R+\overline{f_S^L}
{{{\bi \Phi}_D^{0}}^*}^T{\bi f}_D^R)$ 
terms in our model are extended as 
$\Gamma_{D3}^{d,l}\overline{{\bi f}_D^L}{{\bi \Phi}_D^{0}}f_S^R+\Gamma_{D4}^{d,l}\overline{f_S^L}
{{{\bi \Phi}_D^{0}}}^T{\bi f}_D^R$ and
$\Gamma_{D3}^{u,\nu}\overline{{\bi f}_D^L}{{\bi \Phi}_D^{0}}^*f_S^R+\Gamma_{D4}^{u,\nu}\overline{f_S^L}
{{{\bi \Phi}_D^{0}}^*}^T{\bi f}_D^R$.  
This simplification in our model seems to be reasonable from the following consideration. 
First we can recognize that the coupling constants $\Gamma_S^f, \Gamma^f_{D1,3,4}$ are considered 
to have the hierarchy as $\Gamma_S^f\gg\Gamma^f_{D1,3,4}$, because 
$\Gamma_S^f$ represents the coupling strength for the coupling of all $S_3$ singlet fields as 
$f_S^Lf_S^R\Phi_S^0$,  on the other hand,  $\Gamma^f_{D1,3,4}$ do the coupling strengths 
for the coupling of $S_3$ singlet and doublet fields as $f_D^Lf_D^R\Phi_S^0$, $f_D^Lf_S^R\Phi_D^0$ 
or  $f_D^Rf_S^L\Phi_D^0$.  
Second, although there is considered to be a difference between the coupling strengths 
$\Gamma_{D3}$ and $\Gamma_{D4}$ for the couplings as $\overline{{\bi f}_D^L}
{{\bi \Phi}_D^{0}}f_S^R$ and $\overline{f_S^L}{{{\bi \Phi}_D^{0}}}^T{\bi f}_D^R$, the difference is 
considered to be very small compared to $\Gamma^f_{D3}, \Gamma^f_{D4}$, that is, 
$\Gamma^f_{D3}, \Gamma^f_{D4}\gg\Gamma^f_{D3}-\Gamma^f_{D4}$, then we can assume 
that $\Gamma^f_{D3}=\Gamma^f_{D4}$.
\par 
We express the Higgs fields $\Phi_i^0$'s in Eq. (4) by the vacuum expectation values $v_i$'s  
and the physical Higgs fields $H_i$'s  as  
\begin{align}
&\Phi_S^0=\frac{1}{\sqrt{2}}(v_S+H_S), \notag\\
&\Phi_{1}^0=\cos\alpha \Phi_De^{i\phi_1}=\frac{1}{\sqrt{2}}(v_1+H_1)=\cos\alpha\frac{1}{\sqrt{2}}(v_D+H_D)
e^{i\phi_1}, \\ 
&\Phi_{2}^0=\sin\alpha \Phi_De^{i\phi_2}=\frac{1}{\sqrt{2}}(v_2+H_2)=\sin\alpha\frac{1}{\sqrt{2}}(v_D+H_D)
e^{i\phi_2},  \notag
\end{align}
where we set $\Phi_S^0$ to be real because we can always make the phase of $\Phi_S$ zero by a 
rotation in gauge space.
The mass Lagrangian and mass matrices are obtained on the vacuum expectation values of Higgs fields 
$\Phi_i^0$, and are expressed as
\begin{eqnarray}
&&-{\cal L}_D^f=\overline{f^L}M_ff^R +h.c.,\ \ \ f=d, u, l, \nu, \nonumber\\
&&\hspace{0.5cm}M_{d,l}=\left(\begin{array}{ccc}
     \mu_1^{d,l}+\mu_2^{d,l}e^{i\phi_2}&\lambda\mu_2^{d,l}e^{i\phi_1}&\lambda\mu_3^{d,l}
     e^{i\phi_1}\\
     \lambda\mu_2^{d,l}e^{i\phi_1}&\mu_1^{d,l}-\mu_2^{d,l}e^{i\phi_2}&\mu_3^{d,l}e^{i\phi_2}\\
     \lambda\mu_3^{d,l}e^{i\phi_1}&\mu_3^{d,l}e^{i\phi_2}&\mu_0^{d,l}
     \end{array}\right),\\ 
&&\hspace{0.5cm}M_{u,\nu}=\left(\begin{array}{ccc}
     \mu_1^{u,\nu}+\mu_2^{u,\nu}e^{-i\phi_2}&\lambda\mu_2^{u,\nu}e^{-i\phi_1}&\lambda\mu_3^{u,\nu}
     e^{-i\phi_1}\\
     \lambda\mu_2^{u,\nu}e^{-i\phi_1}&\mu_1^{u,\nu}-\mu_2^{u,\nu}e^{-i\phi_2}&\mu_3^{u,\nu}e^{-i\phi_2}\\
     \lambda\mu_3^{u,\nu}e^{-i\phi_1}&\mu_3^{u,\nu}e^{-i\phi_2}&\mu_0^{u,\nu}
     \end{array}\right),   \notag 
\end{eqnarray}
where we used the following parameterizations, 
\begin{equation}
\begin{array}{l}
\mu_0^{f}=\Gamma_S^{f}\dis{\frac{v_S}{\sqrt{2}}},\ \ \mu_1^{f}=\Gamma_{D1}^{f}\dis{\frac{v_S}{\sqrt{2}}}, \\
\mu_2^{f}=\Gamma_{D2}^{f}\dis{\frac{|v_2|}{\sqrt{2}}}=\Gamma_{D2}^{f}\sin\alpha \frac{v_D}{\sqrt{2}},\ \ 
\mu_3^{f}=\Gamma_{D3}^{f}\frac{|v_2|}{\sqrt{2}}=\Gamma_{D3}^{f}\sin{\alpha}\dis{\frac{v_D}{\sqrt{2}}},\\
\lambda=\dis{\frac{|v_1|}{|v_2|}}=\cot\alpha.\ \ 
\end{array}
\end{equation}
For neutrino mass, we assume the very large Majorana masses,  and from these  
Majorana masses one can get the very small neutrino masses through the see-saw mechanism \cite{SEESAW}. 
We assume that the Majorana mass is constructed from only right handed neutrino 
${\bi \nu}_D^R=(\nu_1^R, \nu_2^R)^T$ and $\nu_S^R$ as to be $S_3$ invariant  and then has no Higgs fields 
\cite{KUBO1}, 
\begin{eqnarray}
{\cal L}_M&=&\frac12\Gamma_S^M(\nu^R_S)^TC^{-1}\nu^R_S+\frac{1}{2}\Gamma_D^M({\bi \nu}^R_D)^TC^{-1}
{\bi \nu}^R_D+h.c.\notag\\
&=&\frac12(\nu_R)^TC^{-1}M_M\nu_R+{h.c.},\\
&&M_M=\left(\begin{array}{ccc}
M_M&0&0\\0&M_1&0\\0&0&M_0
\end{array}\right),\notag
\end{eqnarray}
where $C$ is a charge conjugation matrix. 
\par
From the numerical analyses explaining the masses of quarks and KM mixing matrix 
containing the CP-violation effects, we can get the following numerical results for 11 parameters 
$\mu^f_i$, $\lambda$ and $\phi_i$ \cite{TESHIMA2};
\begin{eqnarray}
&&\mu_0^d=4.20\pm0.12{\rm GeV},\ \frac{\mu_1^d}{\mu_0^d}=0.0120\pm0.0030,\ \ 
\frac{\mu_2^d}{\mu_0^d}=-0.0136\pm0.0004,\nonumber\\
&&\hspace{2cm} \frac{\mu_3^d}{\mu_0^d}=\pm(0.0282\pm0.0008),\nonumber\\
&&\mu_0^u=171.3\pm2.3{\rm GeV},\ \frac{\mu_1^u}{\mu_0^u}=0.00369\pm0.00003,
\ \ \frac{\mu_2^u}{\mu_0^u}=-0.00378\pm0.00003,\\
&&\hspace{2cm} \frac{\mu_3^u}{\mu_0^u}=\mp(0.0127\pm0.0007)\ \ \ ({\rm opposite\ sign\ 
to\ that\ of\ the\ ratio}\ {\mu_3^d}/{\mu_0^d}) ,\nonumber\\
&&\lambda=0.207\pm0.004,\ \ \phi_1=-(74.9\pm0.8)^\circ, \ \ 
\phi_2=(0.74\pm0.31)^\circ.\nonumber
\end{eqnarray} 
From the numerical analysis of charged lepton masses and neutrino mixing , we can get the  
following numerical results for 10 parameters $\mu^f_i$ and $M_i$ \cite{TESHIMA2};
\begin{eqnarray}
&&\mu_0^l=1776.84\pm0.17 {\rm MeV},\ \ \frac{\mu^{l}_1}{\mu_{0}^l}=0.0308\pm0.0007,\ \
\frac{\mu^{l}_2}{\mu_{0}^l}=-(0.0307\pm0.0017),\nonumber  \\
&&\hspace{2cm}\frac{\mu^{l}_3}{\mu_{0}^l}=-0.0233\sim0.0233,\nonumber\\
&&\mu^{\nu}_{0}\approx73.3{\rm GeV}, \ \ 
\frac{\mu^{\nu}_1}{\mu^{\nu}_0}=0.035\sim0.038,\ \ \frac{\mu^{\nu}_2}{\mu^{\nu}_0}
=-0.001\sim-0.007,\\
&&\hspace{2cm}\frac{\mu^{\nu}_3}{\mu^{\nu}_0}=\pm(0.005\sim0.023),\ \notag  \\
&&M_1\approx1.6\times 10^{11}{\rm GeV},\ \ \ M_0\approx 10^{14}{\rm GeV}.\notag 
\end{eqnarray}
From these numerical analyses, we can confirm that $\dis{\frac{\mu_1^f}{\mu_0^f}=\frac
{\Gamma_{D1}^f}{\Gamma_{S}^f}}\approx O(0.01)$  for all flavors, then there is a hierarchy between 
$\Gamma_S^f$ and $\Gamma^f_{D1}$; $\Gamma_S^f\gg\Gamma^f_{D1}$ as mentioned above. 
The result $\lambda=0.207(\alpha=78.3^{\circ})$ predicts the hierarchy between $|v_1|$ and $|v_2|$. 
In almost literature analyzing the flavor mass and mixing using  $S_3$ symmetry 
\cite{PAKUBASA, BHATTACHARYYA, KUBO1, MA, MONDRAGON}, authors assume that $|v_1|=|v_2|$.  
It should be investigated by the analysis for Higgs potential of $S_3$ symmetry whether there is 
a hierarchy as our result $|v_1|/|v_2|=0.207$ or not as $|v_1|=|v_2|$ assumed by other authors. 
From the result $|\mu_1^f|\sim|\mu_3^f|$ as shown in Eqs. (9), (10) and $|\Gamma^f_{D1}|
\sim|\Gamma^f_{D3}|$, which may be considered to be suitable because $|\Gamma^f_{D1}|$ 
and  $|\Gamma^f_{D3}|$ are coupling sterngths for $f_Df_D\Phi_S$ and $f_Df_S\Phi_D$, respectively,  
it is recognized that there is an equality of magnitude for $v_S$ and $v_D$; $v_S\approx v_D$, 
by observing that  $v_S\approx |v_2| =\sin\alpha v_D$ and $\sin\alpha=0.98$.  
Thus, from the quark and lepton mass and mixing analysis \cite{TESHIMA2}, the vacuum expectation 
values and phases of Higgs fields are restricted as 
\begin{align}
v_S\approx v_D, \ \ \ \lambda=\frac{|v_1|}{|v_2|}=\cot\alpha=0.207(\alpha=78.3^\circ), \ \ \ \phi_1
=-74.9^\circ, \ \ \ \phi_2=0.74^\circ.
\end{align}
The purpose of our present work is to investigate whether these results for Higgs fields 
are confirmed or not in $S_3$ invariant Higgs potential.    
\section{$S_3$ invariant Higgs potential } 
The most general $S_3$ invariant Higgs potential composed of quadratic and quartic terms of 
Higgs fields is the following form \cite{PAKUBASA, DERMAN, BRANCO, KOIDE, KUBO2, 
BHATTACHARYYA},    
\begin{align}
V&=-\mu_D^2(\Phi_1^{0\dagger} \Phi_1^0+\Phi_2^{0\dagger} \Phi_2^0)-\mu_S^2\Phi_S^{0\dagger} 
\Phi_S^0\notag\\
&+A(\Phi_S^{0\dagger} \Phi_S^0)^2+B(\Phi_S^{0\dagger} \Phi_S^0)(\Phi_1^{0\dagger} \Phi_1^0+
\Phi_2^{0\dagger} \Phi_2^0)+C(\Phi_1^{0\dagger} \Phi_1^0+\Phi_2^{0\dagger} \Phi_2^0)^2\notag\\
&+D(\Phi_1^{0\dagger} \Phi_2^0-\Phi_2^{0\dagger} \Phi_1^0)^2+E[\Phi_S^{0\dagger}\{\Phi_1^0
(\Phi_1^{0\dagger} \Phi_2^0+\Phi_2^{0\dagger} \Phi_1^0)+\Phi_2(\Phi_1^{0\dagger} \Phi_1^0-
\Phi_2^{0\dagger} \Phi_2^0)\}+h.c.]\notag\\
&+F\{(\Phi_S^{0\dagger} \Phi_1^0)(\Phi_1^{0\dagger} \Phi_S^0)+(\Phi_S^{0\dagger} \Phi_2^0)
(\Phi_2^{0\dagger} \Phi_S^0)\}+F'\{(\Phi_S^{0\dagger} \Phi_1^0)^2+(\Phi_S^{0\dagger} 
\Phi_2^0)^2+h.c.\}\notag\\
&+G\{(\Phi_1^{0\dagger} \Phi_1^0-\Phi_2^{0\dagger} \Phi_2^0)^2+(\Phi_1^{0\dagger} \Phi_2^0+
\Phi_2^{0\dagger} \Phi_1^0)^2\},  
\end{align}
where we disregarded the charged Higgs part $\Phi_i^+$ in gauge Higgs doublets $\Phi_i=
^t \!\!(\Phi_i^+,  \Phi_i^0)$, because we do not consider the effects induced from these charged 
Higgs $\Phi_i^+$, in this analysis.    
In present our analysis, we assume that the coupling constant $E$ describing the strength of the 
coupling between $\Phi_S^0$ and $(\Phi_D^0)^3$ is negligible small,  because all other quartic couplings 
are composed of the pairs $\Phi_S^{0\dagger}\Phi_S^0$ and $\Phi_D^{0\dagger}\Phi_D^0$. 
Authors of literature \cite{PAKUBASA} assumed that the potential is symmetric under the reflection 
$R: \Phi_S^0\to-\Phi_S^0$, and then they settled $E=0$. 
\par   
Using the parameterization Eq. (5),  the potential in Eq. (12) on the vacuum expectation values of Higgs 
fields can be expressed as 
\begin{align}
V=&-\frac12{\mu_S}^2{v_S}^2-\frac12{\mu_D}^2{v_D}^2+\frac14 A{v_S}^4+\frac14B'{v_S}^2{v_D}^2+\frac14
C'{v_D}^4,\notag\\
&B'=B+F+2F'(\cos^2\alpha\cos2\phi_1+\sin^2\alpha\cos2\phi_2),\notag\\
&C'=C+G-(D+G)\sin^22\alpha\sin^2(\phi_1-\phi_2).
\end{align}
From this, the following stationary conditions are obtained,
\begin{align}
\frac{\partial V}{\partial \alpha}=&{v_D}^2\sin2\alpha\left\{\frac12{v_S}^2F'(-\cos2\phi_1+\cos2\phi_2)
-v_D^2(D+G)\cos2\alpha\sin^2(\phi_1-\phi_2)\right\}=0,\\
\frac{\partial V}{\partial \phi_1}=&{v_D}^2\left\{-{v_S}^2F'\cos^2\alpha\sin2\phi_1
-\frac14v_D^2(D+G)\sin^22\alpha\sin2(\phi_1-\phi_2)\right\}=0,\\
\frac{\partial V}{\partial \phi_2}=&{v_D}^2\left\{-{v_S}^2F'\sin^2\alpha\sin2\phi_2
+\frac14v_D^2(D+G)\sin^22\alpha\sin2(\phi_1-\phi_2)\right\}=0,\\
\frac{\partial V}{\partial v_S}=&v_S\left({-\mu_S}^2+A{v_S}^2+\frac12B'{v_D}^2\right)=0,\\
\frac{\partial V}{\partial v_D}=&v_D\left(-{\mu_D}^2+C'{v_D}^2+\frac12B'{v_S}^2\right)=0.
\end{align} 
From Eqs. (15) and (16), we can obtain a relation for $\alpha$, $\phi_1$ and $\phi_2$, 
\begin{align}
\frac{\cos^2\alpha}{\sin^2\alpha}=-\frac{\sin2\phi_2}{\sin2\phi_1}.
\end{align}
Using Eqs. (15) and (16),  the Eq. (14) is satisfied automatically,  then the constraints (14), (15) and  
(16) give the only one relation (19) independent of the coupling constants $D$, $G$, $F'$. 
Then this relation is the first restriction obtained from the $S_3$ invariant Higgs potential.
This relation can be satisfied exactly by the result (11) obtained in numerical analysis of 
quark/lepton mass and mixing \cite{TESHIMA2}. 
In fact using the numerical result (11), the left and right hand sides of relation (19) are given 
as follows, 
\begin{align}
\frac{\cos^2\alpha}{\sin^2\alpha}=(0.207)^2=0.043,\ \ \ 
-\frac{\sin2\phi_2}{\sin2\phi_1}=-\frac{\sin(2\times 0.74^\circ)}{\sin(2\times(-74.9^\circ))}=
0.051. \notag
\end{align} 
Thus the $S_3$ invariant Higgs potential could produce the relation between Cabibbo angle 
$\approx\lambda=\cot\alpha$ and the CP violation phases $\phi_1$ and $\phi_2$  
which were decided  by the quark/lepton mass and mixing through the $S_3$ invariant Yukawa 
interaction.
The authors \cite{PAKUBASA} assuming $\cot\alpha=1$ that is, $|v_1|=|v_2|$ , has settled the 
angles  $\phi_1$, $\phi_2$ as $\phi_1+\phi_2=0$ obtained from the Eq. (19),  and analyzed the 
lepton mass and mixing.  
\par
From Eqs. (17) and (18), the values of $v_S$ and $v_D$ are obtained as 
\begin{align}
{v_S}^2=\frac{4C'{\mu_S}^2-2B'{\mu_D}^2}{4AC'-B'^2}, \ \ \
{v_D}^2=\frac{4A{\mu_D}^2-2B'{\mu_S}^2}{4AC'-B'^2},\ \ \ 
4AC'-B'^2>0.  
\end{align} 
The third relation is obtained from the condition minimizing the potential $V$.  
This relation  is satisfied if there is a hierarchy $A$, $C'\gg B'$ 
between these coupling strengths.  
This hierarchy is recognized from the fact that the coupling constants ($A$, $C'$) are the  
strengths for product of pair  ${\Phi_S^0}^\dagger\Phi_S^0$ or 
${\Phi_D^0}^\dagger\Phi_D^0$, on the other hand  the coupling constant $B'$ is the 
strength for product of different representation pair ${\Phi_S^0}^\dagger\Phi_S^0$ and 
${\Phi_D^0}^\dagger\Phi_D^0$.  
Because ${v_S}^2$ and ${v_D}^2$ are positive, then the following relation must be satisfied among 
parameters $A,\ B',\ C',\ \mu_D^2,\ \mu_S^2$, 
\begin{align}
\frac{B'}{2A}<\frac{{\mu_D}^2}{{\mu_S}^2}<\frac{2C'}{B'}.   \notag
\end{align}
And further, from the hierarchy $(A, C')\gg B'$,  the second restriction is obtained, 
\begin{equation}
\frac{\mu_D}{\mu_S}\ {\rm is\ not\ so\ far\ from\ 1}. 
\end{equation} 
From the numerical result $v_S\approx v_D$ obtained in analysis of quark/lepton mass and 
mixing,  and from Eq. (20),  the relation 
\begin{align}
\frac{\mu_D^2}{\mu_S^2}\approx\frac{2C'+B'}{2A+B'}, \notag
\end{align}
is obtained. 
Using the Eq. (21) and above result, we can predict a relation for the coupling 
strengths $A$ and $C'$ as
\begin{equation}
\frac{C'}{A}\ {\rm is\ not\ so\ far\ from\ 1,} 
\end{equation}
this is the third restriction. 
\par
From the Lagrangian for  the coupling between Higgs fields and 
gauge fields,  one takes the relation as 
\begin{align}
\sqrt{{v_S}^2+|v_1|^2+|v_2|^2}=\sqrt{{v_S}^2+{v_D}^2}=\frac{2m_W}{g}=246{\rm GeV},
\end{align} 
where $g$ is the electroweak coupling.  
If we set the assumption $v_S=v_D$, we can get the values for  $v_S$ as 
\begin{align}
v_S=v_D=\frac{1}{\sqrt{2}}\times246{\rm GeV}=174{\rm GeV}. 
\end{align}
Rewriting the $\Phi_i^0$'s in the Higgs potential (12) by the expression of $\Phi_i^0$'s in 
Eq. (5) and regarding the coefficients  of the product of Higgs fields $H_iH_{i'}, (i,i'=S,D)$,  
we obtain the mass matrix for Higgs fields $H_{S,D}$,
\begin{align}
(H_S, H_D)\left(
\begin{array}{cc}2Av_S^2&B'v_Sv_D\\B'v_Sv_D&2C'v_D^2\end{array}\right)
\left(\begin{array}{c}H_S\\H_D\end{array}\right).  
\end{align}
This is diagonalized approximately in the assumption $\dis{\frac{B'^2v_S^2v_D^2}{(Av_S^2-C'v_D^2)^2} 
\ll 1}$ as 
\begin{align}
\left(\begin{array}{c}H_a\\H_b\end{array}\right)&=\left(
\begin{array}{cc}\cos\beta&\sin\beta\\-\sin\beta&\cos\beta\end{array}\right)
\left(\begin{array}{c}H_S\\H_D\end{array}\right),\ \ 
\tan\beta=\frac{B'v_Sv_D}{2(Av_S^2-C'v_D^2)}, \\
&\left\{\begin{array}{l}
m_{H_a}^2\approx 2Av_S^2+\dis{\frac{B'^2v_S^2v_D^2}{2(Av_S^2-C'v_D^2)}\approx 2Av_S^2},\\
m_{H_b}^2\approx 2C'v_D^2-\dis{\frac{B'^2v_S^2v_D^2}{2(Av_S^2-C'v_D^2)}\approx 2C'v_D^2}.
\end{array}\right.\notag
\end{align}
In the following discussion, we pursue an analysis in the assumption  $\tan\beta\ll 1$, then in 
the approximation for sates  $H_a, H_b$ as 
$$
H_a\approx H_S,\ \ H_b\approx H_D.  
$$
\section{$g_{Hff}$ and flavor changing neutral currents(FCNC)}
Introduced the $S_3$ doublet Higgs, flavor changing neutral currents (FCNC) are produced 
in tree level, and strengths of $g_{Hff}$ are changed from the standard model prediction.  
A prediction of the strengths of $g_{Hff}$ would be very important in the present status where the 
Higgs production rate and the branching ratio for these decays are observed in experiments  
\cite{ATLAS, CMS}.       
Coupling strengths of $g_{Hff'}$ and FCNC are obtained from Yukawa interaction (4), 
inserted the $\Phi_i^0$'s containing the physical Higgs fields $H_i$'s expressed in Eq. (5), 
as
\begin{align}
\sum_{i=S,1,2}[g_{H_iff'}]|H_i|=V_{d,l}^{\dagger}\left[
\left(\begin{array}{ccc}
\mu_1^{d,l}&0&0\\0&\mu_1^{d,l}&0\\0&0&\mu_0^{d,l}
\end{array}\right)\frac{H_S}{v_S}
+\left(\begin{array}{ccc}
0&\lambda\mu_2^{d,l}&\lambda\mu_3^{d,l}\\\lambda\mu_2^{d,l}&0&0\\\lambda\mu_3^{d,l}&0&0
\end{array}\right)e^{i\phi_1}\frac{|H_1|}{|v_1|}\right.
\notag\\
\left. +\left(\begin{array}{ccc}
\mu_2^{d,l}&0&0\\0&-\mu_2^{d,l}&\mu_3^{d,l}\\0&\mu_3^{d,l}&0
\end{array}\right)e^{i\phi_2}\frac{|H_2|}{|v_2|} \right]U_{d,l}, \ \ \ (f, f')=(d, s, b) {\ \rm or\ } (e, \mu, \tau), 
\end{align}
\begin{align}
\sum_{i=S,1,2}[g_{H_iff'}]|H_i|=V_u^{\dagger}\left[
\left(\begin{array}{ccc}
\mu_1^u&0&0\\0&\mu_1^u&0\\0&0&\mu_0^u
\end{array}\right)\frac{H_S}{v_S}
+\left(\begin{array}{ccc}
0&\lambda\mu_2^u&\lambda\mu_3^u\\\lambda\mu_2^u&0&0\\\lambda\mu_3^u&0&0
\end{array}\right)e^{-i\phi_1}\frac{|H_1|}{|v_1|}\right.
\notag\\
\left. +\left(\begin{array}{ccc}
\mu_2^u&0&0\\0&-\mu_2^u&\mu_3^u\\0&\mu_3^u&0
\end{array}\right)e^{-i\phi_2}\frac{|H_2|}{|v_2|} \right]U_u, \ \ \ f, f'=u, c, t.
\end{align}
Where, $V_{f}$ and $U_{f}$ are bi-unitary matrices diagonalizing $M_{f}$ in Eq. (6),  as 
\begin{equation}
V_{f}^\dagger M_fU_{f}={\rm diag}[m_{f1}, m_{f2}, m_{f3}], \ \ \ f=d, l, u.  \notag
\end{equation} 
\par
In Eqs.  (27), (28), because  $\mu_0\gg\mu_1, \mu_2, \mu_3$, $[3,3]$  element are scarcely altered 
by diagonalization,  then $\mu_0^f=m_f,\ f=t, b, \tau$. 
Thus, we can get the predictions,   
\begin{equation}
\dis{g_{H_Sff}=\frac{m_f}{v_S}},\ \ g_{H_1ff}=0, \ \  g_{H_2ff}=0,\ \ v_S=174{\rm GeV}, \ \ f=t, b, \tau, 
\end{equation}   
which are compared to the standard model predictions, 
\begin{equation}
\dis{g_{Hff}=\frac{m_f}{v}},\ \ v=246{\rm GeV}, \ \ f=t, b, \tau. 
\end{equation}  
Next, we estimate the coupling strengths of the FCNC in our model using the values Eqs. (9) and (10) 
for parameters obtained in our previous work \cite{TESHIMA2}, and get the following results;
\begin{align}
&[g_{H_Sdd'}]=\left(\begin{array}{ccc}
-0.00028e^{0.3^\circ i}&0.000076e^{92.2^\circ i}&0.00022e^{-74.6^\circ i}\\
-0.000076e^{90.9^\circ i}&0.00030e^{0.1^\circ i}&0.00066e^{-179.7^\circ i}\\
-0.00022e^{-75.5^\circ i}&0.00066e^{-179.3^\circ i}&0.0241e^{-0.0^\circ i}
\end{array}
\right), \notag\\
&[g_{H_1dd'}]=\left(\begin{array}{ccc}
0.000096e^{-150.5^\circ i}&-0.00034e^{106.1^\circ i}&-0.00068e^{-74.0^\circ i}\\
0.00034e^{104.8^\circ i}&0.000088e^{1.4^\circ i}&0.000088e^{-178.6^\circ i}\\
0.00068e^{-74.9^\circ i}&0.000088e^{-178.3^\circ i}&0.000008e^{0.0^\circ i}
\end{array}
\right), \notag\\
&[g_{H_2dd'}]=\left(\begin{array}{ccc}
0.00033e^{1.9^\circ i}&0.000019e^{-151.4^\circ i}&-0.000089e^{-75.6^\circ i}\\
-0.000019e^{-152.7^\circ i}&0.00030e^{-0.2^\circ i}&-0.00070e^{-179.6^\circ i}\\
0.000089e^{-76.5^\circ i}&-0.00070e^{-179.3^\circ i}&0.000040e^{0.0^\circ i}
\end{array}
\right),\ \ \ d, d'=d, s, b, 
\end{align}
\begin{align}
&[g_{H_Suu'}]=\left(\begin{array}{ccc}
-0.0035e^{0.3^\circ i}&0.00082e^{-91.2^\circ i}&-0.0039e^{74.6^\circ i}\\
-0.00082e^{-90.6^\circ i}&0.0037e^{-0.1^\circ i}&-0.0121e^{179.6^\circ i}\\
0.0039e^{74.8^\circ i}&-0.0121e^{179.3^\circ i}&0.9835e^{0.0^\circ i}
\end{array}
\right), \notag\\
&[g_{H_1uu'}]=\left(\begin{array}{ccc}
0.00094e^{150.0^\circ i}&-0.0039e^{-105.3^\circ i}&0.0126e^{74.7^\circ i}\\
0.0039e^{-104.8^\circ i}&0.00087e^{-0.0^\circ i}&-0.0014e^{-180.0^\circ i}\\
-0.0126e^{74.9^\circ i}&-0.0014e^{179.7^\circ i}&0.000066e^{-0.0^\circ i}
\end{array}
\right),\notag\\
&[g_{H_2uu'}]=\left(\begin{array}{ccc}
0.038e^{-1.2^\circ i}&0.00021e^{160.4^\circ i}&0.0014e^{74.5^\circ i}\\
-0.00021e^{161.0^\circ i}&0.0035e^{0.1^\circ i}&0.0127e^{179.6^\circ i}\\
-0.0014e^{74.7^\circ i}&0.0127e^{179.3^\circ i}&0.00033e^{-0.0^\circ i}
\end{array}
\right), \ \ \ u, u'=u, c, t,
\end{align}
\begin{align}
&[g_{H_Sll'}]=\left(\begin{array}{ccc}
-0.00031e^{-176.9^\circ i}&0.000065e^{-86.1^\circ i}&0.000074e^{108.3^\circ i}\\
-0.000065e^{90.0^\circ i}&0.00031e^{-0.0^\circ i}&0.00024e^{-179.6^\circ i}\\
-0.000074e^{-75.2^\circ i}&0.00024e^{-179.2^\circ i}&0.0102e^{-0.0^\circ i}
\end{array}
\right), \notag\\
&[g_{H_1ll'}]=\left(\begin{array}{ccc}
0.000070e^{33.7^\circ i}&-0.00032e^{-71.5^\circ i}&-0.00023e^{108.6^\circ i}\\
0.00032e^{104.8\circ i}&0.000068e^{-0.4^\circ i}&0.000026e^{179.6^\circ i}\\
0.00023e^{-74.9^\circ i}&0.000026e^{180.0^\circ i}&0.000002e^{-0.0^\circ i}
\end{array}
\right),\notag\\
&[g_{H_2ll'}]=\left(\begin{array}{ccc}
0.00032e^{-175.5^\circ i}&0.000017e^{12.1^\circ i}&-0.000027e^{107.7^\circ i}\\
-0.000017e^{-171.7^\circ i}&0.00031e^{0.0^\circ i}&-0.00025e^{-179.6^\circ i}\\
0.000027e^{-75.7^\circ i}&-0.00025e^{-179.3^\circ i}&0.000012e^{0.0^\circ i}
\end{array}
\right), \ \ \ l, l'=e, \mu, \tau.
\end{align}
The minus signs in some elements in $g_{H_iff'}$ reflect on the signs determined 
when the phases $\phi_1$ and $\phi_2$ are 0. 
\par
We analyze whether the strengths of FCNC coupling $g_{Hiff'}$ obtained above 
satisfy the experimental constraint or not. 
First we analyze the leptonic FCNC induced processes, 
\begin{align}
&\mu^-\to e^-e^+e^-,\notag\\
&\tau^-\to e^-e^+e^-, \ \  \tau^-\to e^+\mu^-\mu^-, \ \ \tau^-\to\mu^+e^-e^-, 
\ \ \tau^-\to\mu^-\mu^+\mu^-, \\
&\tau^-\to e^-\mu^+\mu^-, \ \ \tau^-\to \mu^-e^+e^-. \notag
\end{align}  
The decay ratio of the FCNC induced process  $\mu^-\to e^-e^+e^-$ (Fig. 1(a)) to 
the process  $\mu^-\to\nu_{\mu}e^-\bar{\nu_e}$ is calculated in neglecting the terms 
$O((m_e/m_{\mu})^2)$ and neutrino mixing for the process $\mu^-\to \nu_{\mu}e^-\bar{\nu_e}$, 
as 
\begin{align}
&\frac{\Gamma(\mu^-\to e^-e^+e^-)}{\Gamma(\mu^-\to \nu_{\mu}e^-\nu_e)}=\frac1{2^4}
\dis{\left|\sum_i{\bar g}_{H_i\mu e}{\bar g}_{H_iee}\left(\frac{m_W}{m_{H_i}}\right)^2\right|^2}\notag\\
&\hspace{1cm}\approx\frac1{2^4}\dis{\left|\sum_i{\bar g}_{H_i\mu e}{\bar g}_{H_iee}\right|^2\left(
\frac{m_W}{m_{H_S}}\right)^4},\ \ \ {\rm for\ } m_{H_S}\approx m_{H_1}=m_{H_2},\\
&\hspace{2cm}{\rm where}\ \ \ \bar{g}_{H_iff'}=\frac{g_{H_iff'}}{g/2\sqrt{2}}.\notag
\end{align}  
For the process $\tau^-\to e^-\mu^+\mu^-$,  which has two processes as shown in Fig. 1(b),  
the decay ratio $\Gamma(\tau^-\to e^-\mu^+\mu^-)/\Gamma(\tau^-\to \nu_{\tau}\mu^-
\nu_\mu)$  is expressed as
\begin{align}
&\frac{\Gamma(\tau^-\to e^-\mu^+\mu^-)}{\Gamma(\tau^-\to \nu_{\tau}\mu^-\nu_\mu)}
\approx \frac1{2^4}\left|\sum_i{\bar g}_{H_i\tau e}{\bar g}_{H_i\mu\mu}+
\sum_i{\bar g}_{H_i\tau\mu}{\bar g}_{H_i\mu e} \right|^2\left(\frac{m_W}{m_{H_S}}\right)^4. 
\end{align}
\begin{figure}
\unitlength 0.1in
\begin{picture}( 18.4000, 11.3000)(  4.0000,-12.8000)
%
\special{pn 13}%
\special{pa 400 600}%
\special{pa 700 600}%
\special{fp}%
\special{sh 1}%
\special{pa 700 600}%
\special{pa 634 580}%
\special{pa 648 600}%
\special{pa 634 620}%
\special{pa 700 600}%
\special{fp}%
%
\special{pn 13}%
\special{pa 700 600}%
\special{pa 1000 600}%
\special{fp}%
%
\special{pn 13}%
\special{pa 1000 600}%
\special{pa 2200 200}%
\special{fp}%
\special{sh 1}%
\special{pa 2200 200}%
\special{pa 2130 202}%
\special{pa 2150 218}%
\special{pa 2144 240}%
\special{pa 2200 200}%
\special{fp}%
%
\special{pn 13}%
\special{pa 1000 600}%
\special{pa 1600 800}%
\special{dt 0.045}%
%
\special{pn 13}%
\special{pa 1600 800}%
\special{pa 2200 800}%
\special{fp}%
\special{sh 1}%
\special{pa 2200 800}%
\special{pa 2134 780}%
\special{pa 2148 800}%
\special{pa 2134 820}%
\special{pa 2200 800}%
\special{fp}%
%
\special{pn 13}%
\special{pa 2000 1200}%
\special{pa 1800 1000}%
\special{fp}%
\special{sh 1}%
\special{pa 1800 1000}%
\special{pa 1834 1062}%
\special{pa 1838 1038}%
\special{pa 1862 1034}%
\special{pa 1800 1000}%
\special{fp}%
%
\special{pn 13}%
\special{pa 1800 1000}%
\special{pa 1600 800}%
\special{fp}%
\put(6.4000,-7.7000){\makebox(0,0)[lb]{$\mu^-$}}%
\put(22.4000,-3.2000){\makebox(0,0)[lb]{$e^-$}}%
\put(22.4000,-8.7000){\makebox(0,0)[lb]{$e^-$}}%
\put(20.3000,-12.7000){\makebox(0,0)[lb]{$e^+$}}%
\put(7.2000,-14.5000){\makebox(0,0)[lb]{(a)\ \ $\mu^-\to e^-e^+e^-$}}%
\put(12.7000,-8.3000){\makebox(0,0)[lb]{$H_i$}}%
%
\special{pn 20}%
\special{sh 1}%
\special{ar 1000 600 10 10 0  6.28318530717959E+0000}%
\special{sh 1}%
\special{ar 1000 600 10 10 0  6.28318530717959E+0000}%
\special{sh 1}%
\special{ar 1000 600 10 10 0  6.28318530717959E+0000}%
%
\special{pn 20}%
\special{sh 1}%
\special{ar 1590 800 10 10 0  6.28318530717959E+0000}%
\special{sh 1}%
\special{ar 1600 800 10 10 0  6.28318530717959E+0000}%
\special{sh 1}%
\special{ar 1600 800 10 10 0  6.28318530717959E+0000}%
%
\special{pn 20}%
\special{sh 1}%
\special{ar 1000 600 10 10 0  6.28318530717959E+0000}%
\special{sh 1}%
\special{ar 1000 600 10 10 0  6.28318530717959E+0000}%
\end{picture}%
\vspace{0.8cm}\\
\input{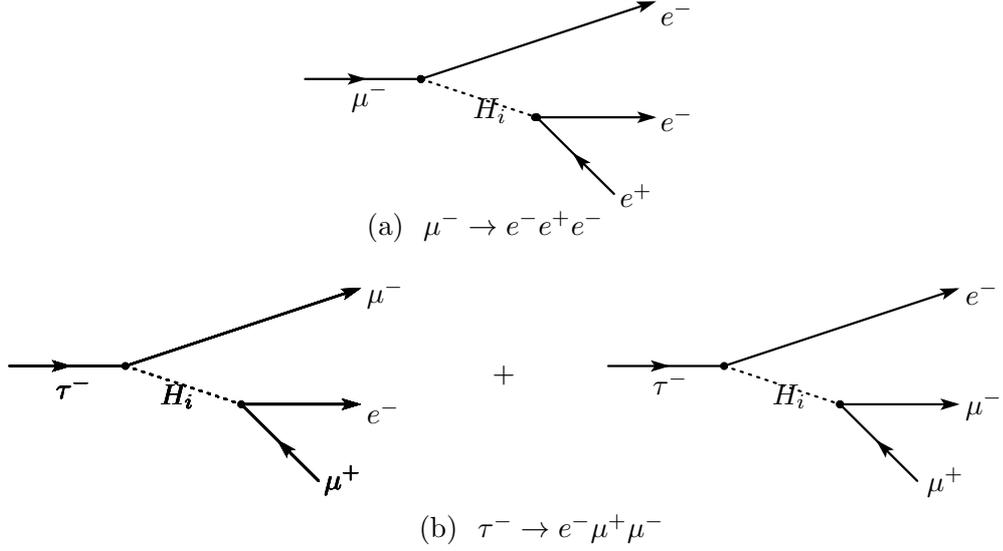}\\
\caption{Leptonic processes   }
\end{figure}
The numerical results for the ratio of these decay widths to ordinary weak decay widths 
are estimated assuming the Higgs mass value is $m_{H_S}=120{\rm GeV}$ and using the experimental 
data \cite{PDG10}. These are tabulated in Table I.  
\begin{table}
\begin{center}
\caption{ Theoretical ratios for the FCNC induced decays and experimental data \cite{PDG10}. }
\begin{tabular}{|l|l|l|}\hline\hline
\ \ \ \ \ Processes        &\ Theoretical ratios    &\ \ \  Experimental data\cite{PDG10} \\ \hline 
$\Gamma(\mu^-\to e^-e^+e^-)/\Gamma(\mu^-\to \nu_{\mu}e^-\bar{\nu_e})$
&$7.5\times10^{-13}$&$<1.0\times10^{-12}$\\  \hline
$\Gamma(\tau^-\to e^-e^+e^-)/\Gamma(\tau^-\to \nu_{\tau}e^-\nu_{e})$
&$5.9\times10^{-13}$&$<1.5\times10^{-7}$\\  \hline
$\Gamma(\tau^-\to \mu^+e^-e^-)/\Gamma(\tau^-\to \nu_{\tau}\mu^-\bar{\nu_{\mu}})$
&$9.5\times10^{-12}$&$<8.6\times10^{-8}$\\  \hline
$\Gamma(\tau^-\to \mu^-e^+e^-)/\Gamma(\tau^-\to \nu_{\tau}\mu^-\nu_{\mu})$
&$1.5\times10^{-11}$&$<1.0\times10^{-7}$\\  \hline
$\Gamma(\tau^-\to e^+\mu^-\mu^-)/\Gamma(\tau^-\to \nu_{\tau}\mu^-\bar{\nu_{\mu}})$
&$1.6\times10^{-13}$&$<9.8\times10^{-8}$\\  \hline
$\Gamma(\tau^-\to e^-\mu^+\mu^-)/\Gamma(\tau^-\to \nu_{\tau}\mu^-\nu_{\mu})$
&$9.1\times10^{-14}$&$<1.6\times10^{-7}$\\  \hline
$\Gamma(\tau^-\to \mu^-\mu^+\mu^-)/\Gamma(\tau^-\to \nu_{\tau}\mu^-\bar{\nu_{\mu}})$
&$2.8\times10^{-15}$&$<1.2\times10^{-7}$\\  \hline
$\Gamma(K^0_L\to e^+e^-)/\Gamma(K^+\to e^+\nu_e)$
&$1.7\times10^{-9}$&$1.4(1\pm{0.67\atop0.44})\times10^{-7}$\\  \hline
$\Gamma(K^0_S\to e^+e^-)/\Gamma(K^+\to e^+\nu_e)$
&$1.3\times10^{-5}$&$<7.9\times10^{-2}$\\  \hline
$\Gamma(K^0_L\to e^\pm \mu^\mp)/\Gamma(K^+\to \mu^+\nu_\mu)$
&$8.9\times10^{-13}$&$<1.8\times10^{-12}$\\  \hline
$\Gamma(K^0_L\to \mu^+ \mu^-)/\Gamma(K^+\to \mu^+\nu_\mu)$
&$1.3\times10^{-16}$&$<2.6\times10^{-9}$\\  \hline
$\Gamma(K^0_S\to \mu^+ \mu^-)/\Gamma(K^+\to \mu^+\nu_\mu)$
&$1.0\times10^{-12}$&$<7.0\times10^{-5}$\\  \hline
$\Gamma(D^0\to e^+e^-)/\Gamma(D^+\to \mu^+\nu_\mu)$
&$1.2\times10^{-9}$&$<5.3\times10^{-4}$\\  \hline
$\Gamma(D^0\to e^\pm\mu^\mp)/\Gamma(D^+\to \mu^+\nu_\mu)$
&$6.6\times10^{-6}$&$<1.7\times10^{-3}$\\  \hline
$\Gamma(D^0\to \mu^+\mu^-)/\Gamma(D^+\to \mu^+\nu_\mu)$
&$1.7\times10^{-9}$&$<9.3\times10^{-4}$\\  \hline
\hline
\end{tabular}
\end{center}
\end{table} 
\par
Next, we analyze the semileptonic decays induced in the FCNC of quarks,  
\begin{align}
K^0_{L,S}\to e^+e^-, \ K^0_{L,S}\to \mu^+\mu^-,\   K^0_{L,S}\to e^\pm\mu^\mp,\  
 D^0\to e^+e^-,\  D^0\to \mu^+\mu^-,\  D^0\to e^\pm\mu^\mp.
\end{align} 
The diagram for these processes, for example,  the process $K^0\to e^+e^-$ is expressed as in Fig. 2. 
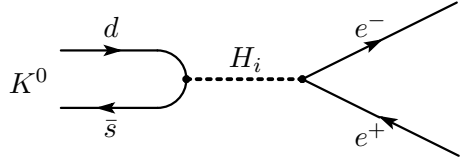
\begin{figure}
\unitlength 0.1in
\begin{picture}( 23.3000,  8.0000)(  1.3000,-13.5000)
%
\special{pn 13}%
\special{pa 400 800}%
\special{pa 700 800}%
\special{fp}%
\special{sh 1}%
\special{pa 700 800}%
\special{pa 634 780}%
\special{pa 648 800}%
\special{pa 634 820}%
\special{pa 700 800}%
\special{fp}%
%
\special{pn 13}%
\special{pa 700 800}%
\special{pa 900 800}%
\special{fp}%
%
\special{pn 13}%
\special{pa 900 1100}%
\special{pa 600 1100}%
\special{fp}%
\special{sh 1}%
\special{pa 600 1100}%
\special{pa 668 1120}%
\special{pa 654 1100}%
\special{pa 668 1080}%
\special{pa 600 1100}%
\special{fp}%
%
\special{pn 13}%
\special{pa 600 1100}%
\special{pa 400 1100}%
\special{fp}%
%
\special{pn 13}%
\special{ar 900 950 150 150  4.7123890 6.2831853}%
\special{ar 900 950 150 150  0.0000000 1.5707963}%
%
\special{pn 20}%
\special{pa 1050 950}%
\special{pa 1650 950}%
\special{dt 0.054}%
\put(6.2000,-7.3000){\makebox(0,0)[lb]{$d$}}%
\put(6.2000,-12.5000){\makebox(0,0)[lb]{$\bar{s}$}}%
\put(12.7000,-9.1000){\makebox(0,0)[lb]{$H_i$}}%
\put(19.2000,-13.0000){\makebox(0,0)[lb]{$e^+$}}%
%
\special{pn 13}%
\special{pa 1650 950}%
\special{pa 2050 750}%
\special{fp}%
\special{sh 1}%
\special{pa 2050 750}%
\special{pa 1982 762}%
\special{pa 2002 774}%
\special{pa 2000 798}%
\special{pa 2050 750}%
\special{fp}%
%
\special{pn 13}%
\special{pa 2050 750}%
\special{pa 2450 550}%
\special{fp}%
%
\special{pn 13}%
\special{pa 2460 1350}%
\special{pa 2060 1150}%
\special{fp}%
\special{sh 1}%
\special{pa 2060 1150}%
\special{pa 2112 1198}%
\special{pa 2108 1174}%
\special{pa 2130 1162}%
\special{pa 2060 1150}%
\special{fp}%
%
\special{pn 13}%
\special{pa 2050 1150}%
\special{pa 1650 950}%
\special{fp}%
\put(19.2000,-7.3000){\makebox(0,0)[lb]{$e^-$}}%
\put(1.3000,-10.2000){\makebox(0,0)[lb]{$K^0$}}%
%
\special{pn 20}%
\special{sh 1}%
\special{ar 1050 950 10 10 0  6.28318530717959E+0000}%
\special{sh 1}%
\special{ar 1050 950 10 10 0  6.28318530717959E+0000}%
%
\special{pn 20}%
\special{sh 1}%
\special{ar 1650 950 10 10 0  6.28318530717959E+0000}%
\special{sh 1}%
\special{ar 1650 950 10 10 0  6.28318530717959E+0000}%
\end{picture}%
\vspace{-0.3cm}\\
\caption{Semileptonic process   }
\end{figure}
We assume the following coupling between scalar current and $K^0$ meson state as
\begin{equation}
J^{(K^0)}(x)=i\sqrt{2}f'_{K}\frac{1}{\sqrt{2p_0V}}e^{ip_\mu x^\mu}, \ \ \ f'_K=m_Kf_K, 
\end{equation} 
where $f_K$ is defined in the coupling between weak current and $K^+$ mesons as 
\begin{equation}
J_\mu^{(K^+)}(x)=i\sqrt{2}f_{K}p_\mu\frac{1}{\sqrt{2p_0V}}e^{ip_\mu x^\mu}, \ \ 
f_K=\tan\theta_C f_\pi, \nonumber
\end{equation}
here, $\theta_C$ is Cabibbo angle, $\tan\theta_C=0.23$,  and $f_\pi=91{\rm MeV}$. 
The ratio of the FCNC induced process, for example  $K^0_L\to e^+e^-$,  to the process 
$K^+\to e^+\nu_e$ is calculated, neglecting the term $O(m_e^2/m_K^2)$ and assuming  
$m_{H_S}\approx m_{H_1}=m_{H_2}$, as 
\begin{align}
\frac{\Gamma(K^0_{L}\to e^+e^-)}{\Gamma(K^+\to e^+\nu_e)}\approx
\frac{m_K^2}{2m_e^2}\left|\frac1{\sqrt{2}}\sum_i\left[{\bar g}_{H_ids}+{\bar g}_{H_isd}\right]
{\bar g}_{H_iee} \right|^2\left(\frac{m_W}{m_{H_S}}\right)^4.
\end{align}
For the ratio of process $K^0_L\to \mu^+\mu^-$ to that of $K^+\to \mu^+\nu_\mu$, 
the kinematics are altered slightly from above as 
\begin{align}
\frac{\Gamma(K^0_{L}\to \mu^+\mu^-)}{\Gamma(K^+\to \mu^+\nu_\mu)}\approx&
\frac{m_K^2(1-\frac{4m_\mu^2}{m_K^2})^{3/2}}{2m_\mu^2(1-
\frac{m_\mu^2}{m_K^2})^2}\left|\frac{1}{\sqrt{2}}\sum_i\left[{\bar g}_{H_ids}+
{\bar g}_{H_isd}\right]{\bar g}_{H_i\mu\mu} \right|^2\left(\frac{m_W}{m_{H_S}}\right)^4. \notag
\end{align} 
For the $K_S^0$ decays, the term $\dis{\left[{\bar g}_{H_ids}+{\bar g}_{H_isd}\right]}$ is replaced as 
$\dis{\left[{\bar g}_{H_ids}-{\bar g}_{H_isd}\right]}$. 
As shown in the numerical result $g_{Hidd'}$ in Eq. (31), $g_{Hids}\approx -g_{Hisd}$,  
thus  the $K^0_L\ {\rm decay}/K^0_S\ {\rm decay}$ ratio becomes very small, that is indicated in 
experimental data \cite{PDG10}. 
The numerical estimations  for the ratios of these decay widths are given by assuming the Higgs 
mass value  $m_{H_S}=120{\rm GeV}$, and  the estimated results and the experimental 
data \cite{PDG10} are shown in Table I. 
Regarding that our estimated results satisfy the experimental data very well,  
we can say that our present $S_3$ invariant model is considered to be fully  realistic
and reasonable model.  
\section{Conclusion}
We constructed our $S_3$ invariant Higgs potential adopting the most general  $S_3$ invariant Higgs 
potential \cite{PAKUBASA, KUBO2, DERMAN, BRANCO, KOIDE, BHATTACHARYYA} and assuming a 
hierarchy between the coupling constants of $S_3$ singlet and doublet Higgs field quartic products.   
We obtained the relation $(|v_1|/|v_2|)^2=\cot^2\alpha=-\sin2\phi_2/\sin2\phi_1$, where 
$v_1$, $v_2$ are vacuum expectation values of $S_3$ doublet Higgs and $\phi_1$, $\phi_2$ are 
phases of  $S_3$ doublet Higgs,  from the stationary condition in our  $S_3$ invariant Higgs potential. 
This relation could be satisfied exactly by the results  $\dis{|v_1|/|v_2|=0.207}$, $\phi_1=
-74.9^\circ$ and $\phi_2=0.74^\circ$ obtained from the quark/lepton mass and mixing analyses 
in the $S_3$ invariant Yukawa interaction \cite{TESHIMA2}. 
Furthermore, we obtained the relation $v_S \sim v_D=\sqrt{v_1^2+v_2^2}=174{\rm GeV}$, which should 
be compared to the standard value $v=246{\rm GeV}$. 
This value affects the coupling strength of Higgs $H_S$ to $f$ quark expressed as $g_{H_Sff}=m_f/v_S$, 
which is altered as by a factor $\sqrt{2}$ to the standard value.  
Introduced the $S_3$ doublet Higgs, FCNC are produced in tree level.
We estimated the branching ratios for rare decays $\mu^-\to e^-e^+e^-$, $\tau^-\to \mu^-\mu^+
\mu^-$, $\cdots$, $K^0_{L,S}\to e^+e^-$, $K^0_{L,S}\to \mu^+\mu^-$, $\cdots$  induced by the FCNC, 
using the values of strength for FCNC estimated in our model.    
The estimated branching ratios satisfy satisfactorily the upper bound obtained from experimental 
data \cite{PDG10}. 
Thus we can say that our present $S_3$ invariant model for Yukawa interaction and Higgs potential is 
a fully realistic and reasonable model.   
\\


\begin{thebibliography}{999}
\bibitem{ATLAS}
ATLAS Collaboration, Report No. ATLAS-CONF-2011-135, 2011, \\
https://twiki.cern.ch/twiki/bin/view/AtlasPublic/AtlasResultsEPS2011.
\bibitem{CMS}
CMS Collaboration, Report No. CMS PAS HIG-11-022, 2011, \\
http://cms.web.cern.ch/cms/News/2011/LP11.
\bibitem{TESHIMA1}
 T. Teshima,  Phys. Rev. {\bf D73}, 045019(1997). 
\bibitem{TESHIMA2}
 T. Teshima and Y. Okumura, Phys. Rev. {\bf D84},  016003(2011). 
\bibitem{SEESAW}
 T.~Yanagida, in {\it Proceedings of the Workshop on the 
    Unified Theories and Baryon Number in the Universe} Tsukuba, {\it 1979}, 
    ed. O.~Sawada and A. Sugamoto, KEK report No.79-18, Tsukuba (1979), p.~95;
    \ 
    M.~Gell-Mann, P. Ramond and R. Slansky, in {\it Supergravity, Proceedings 
    of the Workshop, Stony Brook, New York, 1979}, ed. P.~van Nieuwenhuizen 
    and D.~Freedmann (North-Holland, Amsterdam, 1979), p.~315.
\bibitem{ATMOS}  
 K. Nishikawa, presented at the XI Int. Symp. on Lepton and Photon Interactions 
 at High Energies(Lepton Photon 2003), 
 Fermilab, August, 2003
\bibitem{SOLAR}
Y. Kosio, to appear in the Proceedings of 38th Rencontres de Moriond on Electroweak 
Interactions and Unified Theories, Les Arcs, France, March 15-22, 2003, 
Published in *Tsukuba 2003, Cosmic Ray* 1225-1228. \\
The SNO Collaboration (S. Ahmed {\it et al.}), Phys. Rev. Lett. {\bf 92}, 
181301(2004).\\
The KamLAND Collaboration (K. Eguchi {\it et al.}), Phys. Rev. Lett. {\bf 90}, 
 021802(2003). 
\bibitem{NEUTRINO} M. C. G-Garcia, M. Maltoni, J. Salvado, arXive:1001.4524v3, 
[hep-ph] 15 Apr 2010, JHEP 1004, 056(2010).
\bibitem{PAKUBASA}
 S. Pakvasa and H. Sugawara, Phys. Lett. {\bf 73B}, 61(1978). 
\bibitem{DERMAN}
 E. Derman, Phys. Rev. {\bf D19}, 317 (1979).  
\bibitem{BRANCO}
 G. C. Branco, J. -M. Gerard, and W. Grimus, Phys. Lett. {\bf 136B}, 383, (1984).   
\bibitem{KOIDE} 
Y. Koide, Phys. Rev. {\bf D60}, 077301 (1999). 
\bibitem{KUBO2}
 J. Kubo, H. Okada, and F. Sakamaki, Phys. Rev. {\bf D70}, 036007 (2004).
\bibitem{BHATTACHARYYA}
 G. Bhattacharyya, P. Leser, and H. P\"{a}s, Phys. Rev. {\bf D83}, 011701 (2011).
\bibitem{PDG10}
K. Nakamura {\it et al.} (Particle Data Group), JPG {\bf 37}, 075021 (2010).   
\bibitem{KUBO1}
 J. Kubo, A. Mondrag\'on, M. Mondrag\'on and E. Rodr\'iguez-Jauregui, Prog. Theor. 
 Phys. {\bf 109}, 795(2003).
\bibitem{MA}
 S-L. Chen, M. Frigerio, and E. Ma, Phys. Rev. {\bf D70}, 073008 (2004); 
  Erratum-ibid. {\bf D70}, 079905 (2004).  
\bibitem{MONDRAGON}
 A. Mondrag\'on, M. Mondrag\'on and E. Peinado, Phys. Rev. {\bf D76}, 
 076003(2007).  
\end{thebibliography}
\end{document}